\documentclass{emulateapj}
\usepackage{apjfonts}
\usepackage{graphicx}
\usepackage{psfig}
\usepackage{amssymb}
\newcommand{\begit}{\begin{itemize}}
\newcommand{\enit}{\end{itemize}}
\newcommand{\begen}{\begin{enumerate}}
\newcommand{\enen}{\end{enumerate}}

\newcommand       \be           {\begin{equation}}
\newcommand       \ee           {\end{equation}}
\newcommand       \bea          {\begin{eqnarray}}
\newcommand       \eea          {\end{eqnarray}}

\newcommand       \cm		{\,{\rm cm }}

\newcommand       \yr		{\,{\rm yr }}

\newcommand{\beqa}{\begin{eqnarray}} 
\newcommand{\eeqa}{\end{eqnarray}}

\shorttitle{ON THE WD-WD MERGERS IN TRIPLE SYSTEMS}
\shortauthors{Prodan et. al}

\begin{document}

\title{ON WD-WD MERGERS IN TRIPLE SYSTEMS: THE ROLE OF KOZAI RESONANCE WITH TIDAL FRICTION}

\author{Snezana Prodan\altaffilmark{1}, Norman Murray\altaffilmark{1,2} \& Todd A. Thompson\altaffilmark{3}}

\altaffiltext{1}{Canadian Institute for Theoretical Astrophysics, 60
St.~George Street, University of Toronto, Toronto, ON M5S 3H8, Canada;
sprodan@cita.utoronto.ca} 
\altaffiltext{2}{Canada Research Chair in Astrophysics}
\altaffiltext{3}{Department of Astronomy and Center for Cosmology \& Astro-Particle Physics, The Ohio State University, Columbus, OH 43210, USA}

\begin{abstract}
White dwarf--white dwarf (WD--WD) mergers may lead to type Ia supernovae events. %But since these mergers are driven by gravitational wave radiation the major issue is how to produce enough binaries that are sufficiently tight to merge in a Hubble time so as to reproduce the observed rates of these events.
 \citet{2011ApJ...741...82T} suggested that many such binaries are produced in hierarchical triple systems. The tertiary induces eccentricity oscillations in the inner binary via the Kozai--Lidov mechanism, driving the binary to high eccentricities, and significantly  reducing the gravitational wave merger timescale ($T_{GW}$) over a broad range of parameter space. Here, we investigate the role of  tidal forces in compact hierarchical  systems with $a_{in}=0.05$ AU and $a_{out}=1$ AU, a parameter space relevant for compact systems in the field and in globular cluster.  We show that tidal effects are important in the regime of moderately high initial relative inclination between the inner binary and the outer tertiary. For $85^{o}\leq i_0 \leq90^{o}$ (prograde) and $97^{o}\leq i_0 \leq102^{o}$ (retrograde), tides combine with GW radiation to dramatically decrease $T_{GW}$. In the regime of high inclinations between $91^{o}\leq i_0 \leq96^{o}$, the inner binary likely suffers a direct collision, as in the work of \citet{KatzDong2012} and tidal effects do not play an important role. 
\end{abstract}


%\altaffiltext{1}{Canadian Institute for Theoretical Astrophysics, 60
%St.~George Street, University of Toronto, Toronto, ON M5S 3H8, Canada;
%sprodan@cita.utoronto.ca} 
%\altaffiltext{2}{Canada Research Chair in Astrophysics}

\section{INTRODUCTION}

%Stuff on WD-WD binaries, how these systems are considered SNe Type Ia progenitors, how often these can be in triples, Todd's paper, KCTF and gravitational radiation.

The merger of two white dwarfs (WDs) driven by gravitational wave (GW) radiation has been suggested as a possible mechanism leading to production of type Ia supernovae \citep{1984ApJS...54..335I, 1984ApJ...277..355W, 2011NatCo...2E.350H}. For this mechanism to be observationally relevant, the merger rate has to be comparable to the rate of Ia supernovae events. Even though a large fraction of stars are born as binaries, only a small fraction are tight enough to merge via GW emission within a Hubble time. But if these binaries are in hierarchical triples,  the presence of a tertiary on a highly inclined orbit can induce eccentricity oscillations in the inner binary via secular resonance \citep{1962AJ.....67..591K,1962P&SS....9..719L}. 
%The merger timescale via GW radiation ($T_{GW}$) is a strong function of eccentricity (Peters 1964) and hence the presence of a third body on a highly inclined orbit  pumping the eccentricity of the inner binary to high values will dramatically decrease $T_{GW}$. 
The  GW radiation timescale ($T_{GW }$)  and the time scale associated   with dissipative tides
are both a strong function of eccentricity; hence the presence of a  third body on a highly inclined orbit pumping the eccentricity of the inner binary to high values can dramatically decrease the  merger time scale of the  binary  \citep{ 2002ApJ...578..775B, 2002ApJ...576..894M, 2003ApJ...598..419W, 2012ApJ...757...27A}.

%This was shown explicitly in the work of Blaes et al (2002) for triple systems of supermassive black holes formed in galaxy mergers during structure formation, then in the work of Miller and Hamilton (2002) for four-body interaction of the stellar mass black holes in the globular clusters to accelerate the formation of the intermediate black hole. 

This mechanism was invoked by \citet{2011ApJ...741...82T} to enhance the rate of WD--WD,  NS--WD and NS--NS mergers due to GW radiation. Such mergers are responsible for production of exotica such as Ia supernovae, $\gamma$-ray bursts (GRBs) and other transients. \citet{2011ApJ...741...82T} showed that the GW merger timescale for compact object binaries in triple systems was significantly decreased from that of the binary alone, which allowed for a larger range of the semimajor axes that could lead to a merger in a Hubble time, as well as an increased rate of prompt mergers ($< 10^8$ yr). For detailed discussion on how common these systems are, formation scenarios and rates we refer the reader to \citet{2011ApJ...741...82T}. 

In this paper we explore the role of tides in WD--WD merger events in a compact hierarchical system with $a_{in}=0.05$ AU and $a_{out}= 1$ AU. Such compact systems in the field might be a result of triple-common envelope, as discussed by \citet{1999ApJ...511..324I}. The parameter space we investigate here is relevant for compact systems in the field as well as those in globular clusters. We demonstrate that in the range of high inclinations ($91^{o}\leq i_0 \leq96^{o}$), the outcome of the evolution is a direct collision of the two WDs, in which tidal effects do not play a significant role \citep[see as well][]{KatzDong2012} . Tidal effects do play a significant role in the range of moderately high inclinations ($85^{o}\leq i_0 \leq90^{o}$ and $97^{o}\leq i_0 \leq102^{o}$) where they dramatically decrease $T_{GW}$.

  In Section \ref{sec:dynamics} we describing the
Kozai--Lidov dynamics in the presence of additional forces and dissipation due to tides and gravitational wave radiation. We describe relevant timescales for our dynamical problem. In Section \ref{sec:numerics} we describe the results of numerical integrations of the equations of motion. We discuss our findings in Section \ref{sec:discussion}.

\section{UNDERSTANDING THE DYNAMICS} \label{sec:dynamics}

%
%\bea \label{eq:hamiltonian} %$ 
%H&=&\frac{-3A}{2}\Bigg[-{5\over3}-3\frac{\Ha^{2}}{\La^{2}}+\frac{\G^{2}}{\La^{2}}+5\frac{\Ha^{2}}{\G^{2}}+5\cos2\omega\left(1-\frac{\G^{2}}{\La^{2}}-\frac{\Ha^{2}}{\G^{2}}+\frac{\Ha^{2}}{\La^{2}}\right)\Bigg]\nonumber\\ &&-B\frac{\La}{\G}-k_2C\left(35\frac{\La^{9}}{\G^{9}}-30\frac{\La^{7}}{\G^{7}}+3\frac{\La^{5}}{\G^{5}}\right)
%-k_2D\frac{\La^{3}}{\G^{3}}, 
%\eea %$
%

%\documentclass{aastex}
%\bibliographystyle{apj}
%\begin{document}
%\newcommand       \cm		{\,{\rm cm }}

\begin{table}
\begin{center}
\begin{tabular}{|l|l|l|l|}
\tableline
\multicolumn{3}{|c|}{TABLE 1. System parameters} \\
\tableline
\tableline
Symbol & Definition & Value  \\ \hline
$m_1$ & White dwarf (primary) mass & $0.8 M_{\bigodot}$ \\
$m_2$ & White dwarf (secondary) mass & $0.6 M_{\bigodot}$ \\
$m_3$ & Third companion mass & $1.0 M_{\bigodot}$\\
$a_{in}$ & Inner binary semimajor axis & $0.05 AU$\\
$a_{out}$ & Outer binary semimajor axis & $1 AU$ \\
$e_{in, 0}$ & Inner binary initial eccentricity & $0.1$ \\
$e_{out, 0}$ & Outer binary eccentricity & $0.5$ \\
$i_{init}$ & Initial mutual inclination & $85^{o}-102^{o}$ \\
$\omega_{in, 0}$ & Initial argument of periastron & $0$\\
$\Omega_{in}$ & Longitude of ascending node & $0$\\
$R$ & White dwarf radius & $5\times10^8 \cm$\\
$k_2$ & Tidal Love number & $0.1$\\
$Q$ & Tidal dissipation factor & $10^7$\\
\tableline
\tableline
\end{tabular}
\end{center}
\end{table}

%\bibliography{1820}{}

%\end{document}

\subsection{The Kozai--Lidov mechanism}
%Maybe add some words on this later: basic explanation of how it works, then how GR, TBs, RBs and tidal dissipation can mess around with it, mention how it is used in different contexts (i.e triple stellar systems (my work, todd's, fabrycky, blaes), Galactic center (Fabio's work), planetary context (Yanqin, Smadar, Fabrycky). 

The presence of a third body on a hierarchical orbit around the centre of mass of a binary will affect the orbital elements of the binary on a variety of timescales. The induced changes in the orbital elements of the binary will be particularly striking  if the mutual inclination between the inner and the outer orbit is high. The two orbits will exchange angular momentum, causing both the eccentricity of the inner binary and the mutual inclination to undergo periodic oscillations known as Kozai cycles \citep{1962AJ.....67..591K}. \citet{1962AJ.....67..591K} showed that the mutual inclination required for having Kozai cycles is $i_{crit}\leq i \leq180^{o}-i_{crit}$, where $i_{crit} \approx 39.2^{o}$ is a critical inclination. Kozai cycles result from near -- $1:1$ resonance between the longitude of the periapse $\varpi$ and the longitude of the ascending node $\Omega$ and therefore the condition for Kozai resonance, $\dot{\varpi}-\dot{\Omega}=0$, is fulfilled only for inclinations in the Kozai regime ($i_{crit}\leq i \leq180^{o}-i_{crit}$). For inclinations outside of the Kozai regime, the apsidal line precesses in a prograde sense ($\dot{\varpi}>0$), while the line of nodes precesses in a retrograde sense ($\dot{\Omega}<0$). For prograde orbits ($i\leq90^{o}$) these cycles are out of phase, meaning that when the eccentricity reaches its maximum, the mutual inclination reaches its minimum and vice versa. On the other hand, for retrograde orbits ($i>90^{o}$) these cycles are in phase: both the eccentricity and the mutual inclination reach maximum values simultaneously. When the exact resonance condition is satisfied, the system is trapped in a resonance and librates around the fixed point. The period of a Kozai cycle is significantly longer than either the orbital period of the inner or the outer binary, suggesting the use of the secular approximation. The secular approximation consists of averaging the equations of motion over the orbital periods of the inner and the outer binary. Such averaged equations allow for exchange of angular momentum between the two orbits but not variations in the energy, so that the semimajor axes of both orbits remain unchanged. The relevance of the secular approximation is discussed further in Section \ref{sec:discussion}. The maximum eccentricity attained in Kozai cycles in the absence of additional forces is given by: 
\be %$
e_{max} = \left(1-\frac{5}{3}cos^2i_0\right)^{\frac{1}{2}}\label{eq:emax}
\ee %$
where $i_0$ is the initial mutual inclination between the inner and the outer orbit. Note that equation \ref{eq:emax} is given in the test particle limit where the secondary is treated as a massless particle.

Kozai cycles can be suppressed by other dynamical effects that induce periapse precession in the inner binary. These dynamical effects affect the orientation of the inner orbit in a sense that the two orbit will no longer be in the configuration that allows them to exchange angular momentum for  the same amount of time as when those effects are negligible. As a result, the maximum possible eccentricity attainable by the system is smaller and the critical inclination becomes larger. In other words, near --  $1:1$ resonance condition between the longitude of the periapse $\varpi$ and the longitude of the ascending node $\Omega$ is only satisfied for higher initial inclinations. We take into account the following additional sources of periapse precession: apsidal precession due to tidal and rotational bulges, apsidal precession due to general relativity (GR) in the inner orbit (first order post--Newtonian expansion) and the apsidal precession due to tidal dissipation in the stars in the inner binary, which is negligible in comparison to the other two. Tidal dissipation as well as gravitational wave radiation play a major role in driving the merger of the inner binary and are discussed in more detail later in the paper.

We consider an inner WD--WD binary with semimajor axis $a_{in}$, eccentricity $e_{in}$, argument of periastron $\omega_{in}$ , masses $m_1$ and $m_2$ and two equal radii $R$. The third body with mass $m_3$, semimajor axis $a_{out}$, eccentricity $e_{out}$ and argument of periastron $\omega_{out}$ is on a larger orbit around the center of mass of the inner binary. The mutual inclination between the two orbits is $i$. The mean motion of the inner binary is $n=[G(m_1+m_2)/a^{3}_{in}]^{1/2}$ Fiducial values of the system parameters used throughout this paper are listed in  Table 1. Subscript ``0'' denotes initial values (e.g. $e_{in,0}$). 

The equations for the precession rates due to the Kozai--Lidov mechanism, GR, tidal and rotational bulges raised on both WDs in the quadrupole approximation are \citep{2001ApJ...562.1012E}:

%\bea \label{eq:omega_dots}
%\dot{\omega}_{Kozai}&=&\frac{3}{4}\frac{Gm_3}{a_{out}^3(1-e^{2}_{out})^{\frac{3}{2}}n}\frac{1}{\sqrt{1-e^{2}_{in}}}\Bigg[ 2(1-e^{2}_{in})+5\sin^{2}\omega(e^{2}_{in}-\sin^{2}i)\Bigg] \label{eq:omega_kozai}\\
%\dot{\omega}_{GR}&=&\frac{3(G(m_1+m_2))^{\frac{3}{2}}}{a_{in}^{\frac{5}{2}}c^{2}(1-e^{2}_{in})}\\
%\dot{\omega}_{TB}&=&\frac{15(G(m_1+m_2))^{\frac{1}{2}}}{16a_{in}^{\frac{13}{2}}}\frac{8+12e^{2}_{in}+e^{4}_{in}}{(1-e^{2}_{in})^{5}}\left(\frac{m_{1}}{m_{2}}+\frac{m_{2}}{m_{1}}\right)k_{2}R^{5}\\
%\dot{\omega}_{RB}&=&\frac{(m_1+m_2)^{\frac{1}{2}}}{4G^{\frac{1}{2}}a_{in}^{\frac{7}{2}}(1-e^{2}_{in})^{2}}k_{2}R^{5}\\
% &\times&\sum_{i=1,2}\frac{1}{m_{i}}\Bigg[\left(2\Omega^{2}_{ih}-\Omega^{2}_{ie}-\Omega^{2}_{iq}\right)+2\Omega_{ih}\cot i\left(\Omega_{ie}\sin\omega_{in} +\Omega_{iq}\cos\omega_{in}\right)\Bigg]\\
%\dot{\omega}_{TD}&=&\frac{\cot i}{2nt_{Fi}}\sum_{i=1,2}\left(-\Omega_{ie}\cos\omega_{in}\frac{1+\frac{3}{2}e^2_{in}+\frac{1}{8}e^4_{in}}{(1-e^2_{in})^5}+\Omega_{iq}\sin\omega_{in}\frac{1+\frac{9}{2}e^2_{in}+\frac{5}{8}e^4_{in}}{(1-e^2_{in})^5}\right)
%\eea
%%

\bea \label{eq:omega_dots}
\dot{\omega}_{Kozai}&=&\frac{3}{4}\frac{Gm_3}{a_{out}^3(1-e^{2}_{out})^{\frac{3}{2}}n}\frac{1}{\sqrt{1-e^{2}_{in}}}\times\\ \nonumber
&&\Bigg[ 2(1-e^{2}_{in})+5\sin^{2}\omega_{in}(e^{2}_{in}-\sin^{2}i)\Bigg] \label{eq:omega_kozai}\\
\dot{\omega}_{GR}&=&\frac{3(G(m_1+m_2))^{\frac{3}{2}}}{a_{in}^{\frac{5}{2}}c^{2}(1-e^{2}_{in})}\\
\dot{\omega}_{TB}&=&\frac{15(G(m_1+m_2))^{\frac{1}{2}}}{16a_{in}^{\frac{13}{2}}}\frac{8+12e^{2}_{in}+e^{4}_{in}}{(1-e^{2}_{in})^{5}}\times\\
&&\left(\frac{m_{1}}{m_{2}}+\frac{m_{2}}{m_{1}}\right)k_{2}R^{5}\nonumber\\
\dot{\omega}_{RB}&=&\frac{(m_1+m_2)^{\frac{1}{2}}}{4G^{\frac{1}{2}}a_{in}^{\frac{7}{2}}(1-e^{2}_{in})^{2}}k_{2}R^{5}\times\\ \nonumber
 &&\sum_{i=1,2}\frac{1}{m_{i}}\Bigg[\left(2\Omega^{2}_{ih}-\Omega^{2}_{ie}-\Omega^{2}_{iq}\right)\Bigg. \nonumber\\ +  &&\Bigg. 2\Omega_{ih}\cot i\left(\Omega_{ie}\sin\omega_{in} +\Omega_{iq}\cos\omega_{in}\right)\Bigg]\nonumber\\
\dot{\omega}_{TD}&=&\frac{\cot i}{2nt_{Fi}}\sum_{i=1,2}\left(-\Omega_{ie}\cos\omega_{in}\frac{1+\frac{3}{2}e^2_{in}+\frac{1}{8}e^4_{in}}{(1-e^2_{in})^5}\right.\\ \nonumber\\&&+\left. \Omega_{iq}\sin\omega_{in}\frac{1+\frac{9}{2}e^2_{in}+\frac{5}{8}e^4_{in}}{(1-e^2_{in})^5}\right)\nonumber
\eea

where $t_{F1}=\frac{1}{6}\left(\frac{a_{in}}{R}\right)^{5}\frac{1}{n}\left(\frac{m_{2}}{m_{1}}\right)\frac{Q}{k_{2}}$ is the tidal friction time scale for the star with mass $m_1$. A similar expression with two indices $1$ and $2$ swapped holds for the tidal friction time scale $t_{F2}$ for the star  with mass $m_2$. The three components of the spin, $\Omega_i$, are  the projection along the Laplace-Runge-Lenz vector, pointing along the apsidal line from the WD secondary at the apoapse toward
the WD primary, denoted by $\Omega_e$, along the total angular
momentum vector, $\Omega_h$, and their cross product, denoted by $\Omega_q$.

As equation \ref{eq:omega_kozai} shows, the term driving Kozai cycles can be either positive or negative depending on the value of $\sin i$. On the other hand, the terms driving the precession due to GR and the tidal bulge are always positive and therefore tend to promote periapse precession. The  effect of these two terms is to lower the maximum eccentricity attainable by the system, while the critical inclination increases \citep[see their Figure 3]{2001ApJ...562.1012E, 2002ApJ...576..894M, 2003ApJ...589..605W, 2007ApJ...669.1298F}. The term induced by the rotational bulge may have either positive or negative value. We assume that initially the system is tidally locked and that the spins are aligned, so this term tends to increase the rate of precession and hence suppress Kozai cycles. Precession due to both rotational and tidal bulges are parametrized by the tidal Love number $k_2$ which is a dimensionless constant that relates the mass of the multipole moment created by tidal forces on the spherical body to the gravitational tidal field in which that same body is immersed. Furthermore, $k_2$ encodes information on the internal structure of the body in question\footnote{The apsidal precession constant, which is a factor of two smaller than the tidal Love number, but which we do not utilize, is often denoted by $k_2$ as well.}. 

Tidal dissipation in the stars in the inner binary, due to either an eccentric orbit or to asynchronous rotations, becomes important when the separation between the stars is of order of a few stellar radii \citep{1979A&A....77..145M, 2001ApJ...562.1012E}. During the phases of high eccentricity the periapse distance may become sufficiently small as to lead to strong tidal dissipation. During these phases of high eccentricity the tidal dissipation will drain energy from the orbit, but not angular momentum. The energy loss results in a reduction of the semimajor axis and therefore enhances the rate of dissipation. Since the angular momentum remains conserved during this process, the eccentricity is damped as well until the orbit eventually circularizes and the system settles at a separation of only a few stellar radii. Tidal dissipation is parametrized by the tidal dissipation factor $Q$, defined as the ratio of the energy stored in the tidal bulge to the energy dissipated per orbit.  

Another source of dissipation in WD--WD binaries is gravitational wave radiation which drains both energy and angular momentum from the orbit. Like tidal dissipation, it tends to shrink and circularize the orbit. During the Kozai cycles, if the amplitude of the eccentricity oscillations is sufficiently large, the GW radiation becomes much stronger than in the circular case, leading to mergers on timescales much shorter than a single Hubble time, $T_{Hubble}$ \citep{2002ApJ...578..775B, 2002ApJ...576..894M, 2011ApJ...741...82T,  2012ApJ...757...27A}.  As the inner eccentricity reaches values close to $1$, dissipation due to tides and GW radiation become comparable and neither can be neglected, as we discuss in more detail in next section ( see figure \ref{Fig:timescales}).

\subsection{Timescales}

The GW merger timescale in the limit of high eccentricity is \citep{1964PhRv..136.1224P}:
%\bea\label{eq: Tgw}
%T_{GW} &=& \frac{3}{85}\frac{a_{in}}{c}\left(\frac{a_{in}^3c^6}{G^3m_1m_2 M}\right) (1-e_{in}^2)^{\frac{7}{2}}\nonumber\\
%               &\simeq& 5.4\times10^{12}~yr\left(\frac{0.672M_{\bigodot}^3}{m_1m_2M}\right)\left(\frac{a_{in}}{0.05AU}\right)^4(1-e_{in}^2)^{\frac{7}{2}},
%\eea

\bea\label{eq: Tgw}
T_{GW} &=& \frac{3}{85}\frac{a_{in}}{c}\left(\frac{a_{in}^3c^6}{G^3m_1m_2 M}\right) (1-e_{in}^2)^{\frac{7}{2}}\\
               &\simeq& 5.4\times10^{12}~yr\left(\frac{0.672M_{\bigodot}^3}{m_1m_2M}\right)\left(\frac{a_{in}}{0.05AU}\right)^4(1-e_{in}^2)^{\frac{7}{2}},\nonumber
\eea
where $M=m_1+m_2$.   As seen from eqn \ref{eq: Tgw}, in the absence of a third body $T_{GW}$ is greater than the Hubble time, $T_{Hubble}\simeq 14~Gyr$, for $a_{in} > 0.01AU$.  

As discussed in the previous section, GR precession tends to promote periapse precession and therefore suppress Kozai cycles. In general, GR precession decreases the maximum possible eccentricity attainable by the binary at a fixed initial inclination and increases the critical inclination required for undergoing Kozai cycles \citep{2002ApJ...578..775B, 2007ApJ...669.1298F,2012ApJ...747....4P}. The timescale for GR precession is:
%\bea\label{eq:Tgr}
%T_{GR}&=&\frac{1}{3}\frac{a_{in}}{c}\left(\frac{a_{in}c^2}{GM}\right)^{\frac{3}{2}}(1-e_{in}^2) \nonumber\\
%&\simeq&1.8\times10^3~yr\left(\frac{1.4 M_{\bigodot}}{M}\right)^{\frac{3}{2}}\left(\frac{a_{in}}{0.05AU}\right)^{\frac{5}{2}}(1-e_{in}^2).
%\eea  
\bea\label{eq:Tgr}
T_{GR}&=&\frac{1}{3}\frac{a_{in}}{c}\left(\frac{a_{in}c^2}{GM}\right)^{\frac{3}{2}}(1-e_{in}^2) \\
&\simeq&1.8\times10^3~yr\left(\frac{1.4 M_{\bigodot}}{M}\right)^{\frac{3}{2}}\left(\frac{a_{in}}{0.05AU}\right)^{\frac{5}{2}}(1-e_{in}^2).\nonumber
\eea  

 We consider the effects of tidal forces, where both rotational and tidal bulges tend to suppress Kozai cycles in a similar manner as GR precession (Equation \ref{eq:Tgr}). The timescale for precession induced by the tidal bulge is given by \citep{2012ApJ...747....4P, 2007ApJ...669.1298F}:
% \bea\label{eq:Ttb}
% T_{TB}&=&\frac{16}{15k_2}\left(\frac{a_{in}}{R}\right)^5\left(\frac{a_{in}^3}{GM}\right)^{\frac{1}{2}}\left(\frac{m_2}{m_1}+\frac{m_1}{m_2}\right)^{-1}\frac{(1-e_{in}^2)^5}{8+12e_{in}^2+e_{in}^4}\nonumber\\
% &\simeq&5.9\times10^{13}~yr\left(\frac{a_{in}}{0.05AU}\right)^{\frac{13}{2}}\left(\frac{5\times10^8\cm}{R}\right)^5\left(\frac{1.4M_{\bigodot}}{M}\right)^{\frac{1}{2}}\frac{(1-e_{in}^2)^5}{8+12e_{in}^2+e_{in}^4},
% \eea
  \bea\label{eq:Ttb}
 T_{TB}&=&\frac{16}{15k_2}\left(\frac{a_{in}}{R}\right)^5\left(\frac{a_{in}^3}{GM}\right)^{\frac{1}{2}}\left(\frac{m_2}{m_1}+\frac{m_1}{m_2}\right)^{-1}\frac{(1-e_{in}^2)^5}{8+12e_{in}^2+e_{in}^4}\nonumber\\
 &\simeq&5.9\times10^{13}~yr\left(\frac{a_{in}}{0.05AU}\right)^{\frac{13}{2}}\left(\frac{5\times10^8\cm}{R}\right)^5\times\nonumber \\ 
 &&\left(\frac{1.4M_{\bigodot}}{M}\right)^{\frac{1}{2}}\frac{(1-e_{in}^2)^5}{8+12e_{in}^2+e_{in}^4},
 \eea
 where $R$ is the radius of the white dwarf and $k_2=0.1$. The Kozai timescale is \citep{1997AJ....113.1915I, 1997Natur.386..254H}: 
% \bea\label{eq:Tkoz}
% T_{Kozai}&=&\frac{4}{3}\left(\frac{a_{in}^3M}{Gm_3^2}\right)^{\frac{1}{2}}\left(\frac{a_{out}}{a_{in}}\right)^3(1-e_{out}^2)^{\frac{1}{2}}\nonumber\\
% &\simeq& 22~yr\left(\frac{a_{in}}{0.05AU}\right)^{\frac{3}{2}}\left(\frac{M}{1.4M_{\bigodot}}\right)^{\frac{1}{2}}\left(\frac{M_{\bigodot}}{m_3}\right)\left(\frac{a_{out}/a_{in}}{20}\right)^3(1-e_{out}^2)^{\frac{1}{2}}.
% \eea
 \bea\label{eq:Tkoz}
 T_{Kozai}&=&\frac{4}{3}\left(\frac{a_{in}^3M}{Gm_3^2}\right)^{\frac{1}{2}}\left(\frac{a_{out}}{a_{in}}\right)^3(1-e_{out}^2)^{\frac{1}{2}}\\
 &\simeq& 22~yr\left(\frac{a_{in}}{0.05AU}\right)^{\frac{3}{2}}\left(\frac{M}{1.4M_{\bigodot}}\right)^{\frac{1}{2}}\left(\frac{M_{\bigodot}}{m_3}\right)\nonumber\\
 &&\times\left(\frac{a_{out}/a_{in}}{20}\right)^3(1-e_{out}^2)^{\frac{1}{2}}.\nonumber\
 \eea
The condition for the inner binary to undergo Kozai oscillation is that the Kozai timescale is shorter than the timescale of any of the suppressing effects. Furthermore, the Kozai timescale is strongly dependent on the ratio of the inner and the outer binary semimajor axis, which sets a limit on the maximum allowed $a_{out}$, beyond which the Kozai oscillations are ineffective. During the evolution, depending on the type of stars in the inner binary (i.e. compact objects or main sequence stars) and tightness of its orbit, dissipative effects due to gravitational wave radiation or tides may be significant. Either of the two may lead to shrinkage and/or circularization of the inner orbit, which increases the ratio of the inner and outer semimajor axis and thus reduces the effectiveness of Kozai oscillations. Consequently,  the Kozai timescale may become larger than either $T_{GR}$ or $T_{TB}$ and at this point the evolution of the system toward merger would be completely dominated by non-Kozai effects. 

To emphasize the importance of Kozai oscillations for rapid mergers of compact objects and considering only the presence of a third body, \citet{2011ApJ...741...82T} gives the following order-of-magnitude estimate of the merger time: 
%\bea\label{eq:Tmerge}
%T_{merge}&\thicksim& \frac{25}{153}\frac{a_{in}}{c}\left(\frac{a_{in}^3c^6}{G^3m_1m_2M}\right)cos^6i\nonumber\\
%&\thicksim&4.5\times10^4~yr\left(\frac{0.672M_{\bigodot}^3}{m_1m_2M}\right)\left(\frac{a_{in}}{0.05AU}\right)^4\left(\frac{cos i}{cos(88^{o})}\right)^6,
%\eea
\bea\label{eq:Tmerge}
T_{merge}&\thicksim& \frac{25}{153}\frac{a_{in}}{c}\left(\frac{a_{in}^3c^6}{G^3m_1m_2M}\right)cos^6i\\
&\thicksim&4.5\times10^4~yr\left(\frac{0.672M_{\bigodot}^3}{m_1m_2M}\right)\left(\frac{a_{in}}{0.05AU}\right)^4\left(\frac{cos i}{cos(88^{o})}\right)^6,\nonumber
\eea
which shows a strong dependance on mutual inclination. Equation \ref{eq:Tmerge} is only valid in the Kozai regime, $39^{o}\leq i \leq 141^{o}$ and fails to account for additional sources of apsidal precession such as GR precession and precession due to the tidal and/or rotational bulge. As pointed out by \citet{2011ApJ...741...82T}, this expression significantly underestimates the merger time in certain regions of  parameter space and hence should be used only to obtain a rough estimate. For detailed discussion we refer the reader to the Section 4 of \citet{2011ApJ...741...82T}.

The timescale for the semimajor axis to decay due to either tidal dissipation or GW radiation is strongly dependent on the eccentricity: 

%\bea  \label{eq:a_dots}
%     \tau_a|_{TD}= \left(\frac{a}{\dot{a}}\right)_{TD}&=&-\frac{9}{2}t_{F}\frac{1-e^{2}}{e^{2}}\left[\frac{1+\frac{15}{4}e^{2}+\frac{15}{8}e^{4}+\frac{5}{64}e^{6}}{(1-e^{2})^{\frac{13}{2}}}-\frac{11}{18}\frac{1+\frac{3}{2}e^{2}+\frac{1}{8}e^{4}}{(1-e^{2})^{5}}\right]^{-1}\\
%          \tau_a|_{GW}= \left(\frac{a}{\dot{a}}\right)_{GW}&=&-\frac{5c^5a_{in}^3(1-e_{in}^2)^{\frac{7}{2}}}{64Gm_1m_2(m_1+m_2)}\left[1+ \frac{73}{24} e_{in}^2+\frac{37}{96} e_{in}^4\right]^{-1}
%%     \dot{a}_{MT}&=&-\frac{2}{3} a\frac{\dot{m}_{2}}{m_2} 
%\eea

\bea  \label{eq:a_dots}
     \tau_a|_{TD}= \left(\frac{a}{\dot{a}}\right)_{TD}&=&-\frac{9}{2}t_{F}\frac{1-e^{2}}{e^{2}}\times\\
     &&\left[\frac{1+\frac{15}{4}e^{2}+\frac{15}{8}e^{4}+\frac{5}{64}e^{6}}{(1-e^{2})^{\frac{13}{2}}}-\frac{11}{18}\frac{1+\frac{3}{2}e^{2}+\frac{1}{8}e^{4}}{(1-e^{2})^{5}}\right]^{-1}\nonumber\\ 
          \tau_a|_{GW}= \left(\frac{a}{\dot{a}}\right)_{GW}&=&-\frac{5c^5a_{in}^3(1-e_{in}^2)^{\frac{7}{2}}}{64Gm_1m_2(m_1+m_2)}\left[1+ \frac{73}{24} e_{in}^2+\frac{37}{96} e_{in}^4\right]^{-1}
%     \dot{a}_{MT}&=&-\frac{2}{3} a\frac{\dot{m}_{2}}{m_2} 
\eea

\begin{figure}
\epsscale{0.95} 
\plotone{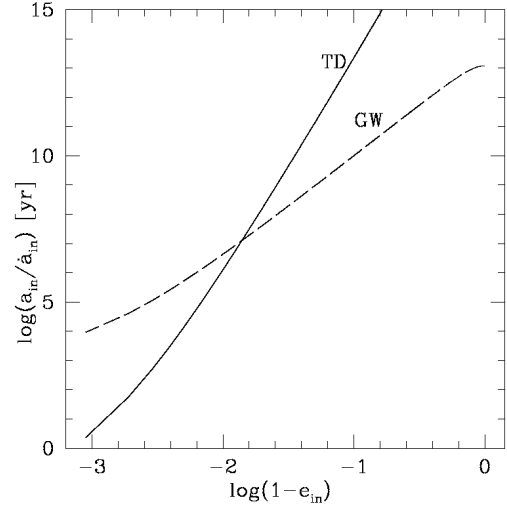}
\caption{
  The timescale for semimajor axis to decay $\tau_a=a_{in}/\dot{a}$ as a function of the eccentricity of the inner binary  for our fiducial model. The solid line represents $\tau_a$ due to tidal dissipation (TD) while the dashed line shows $\tau_a$ due to gravitational wave radiation (GW). As the eccentricity of the inner binary exceeds $e_{in}\approx0.95$ the decay rate of the semimajor axis due to tidal dissipation becomes comparable to gravitational wave radiation, and therefore cannot be neglected. \label{Fig:timescales}}
\end{figure}

As figure \ref{Fig:timescales} shows, during the phases of the Kozai cycle where   $e_{in} \sim 1$, $\tau_a|_{TD}\approx \tau_a|_{GW}$ and hence tidal dissipation can not be ignored. We demonstrate in Section \ref{sec:moderate} that the merger timescale  of the WD--WD binary in the regime of moderately high inclinations ($85^{o}\leq i_0 \leq90^{o}$ and $97^{o}\leq i_0 \leq102^{o}$) is shorter by at least an order of magnitude when tidal dissipation is taken into consideration than when gravitational wave radiation alone is accounted for. 

On the other hand, in the regime of highly inclined orbits ($91^{o}\leq i_0 \leq96^{o}$), neither of the two sources of dissipation affects the merger time of the inner binary since it occurs at the first Kozai maximum.  The timescale to reach Kozai maximum is too short for tides or gravitational wave radiation to kick in (for details see Section \ref{sec:high}). Such collisions at the first Kozai maximum were noted in the work of \citet{2011ApJ...741...82T} as well, and they were shown to explicitly occur in the work of \citet{KatzDong2012}.

%
%%\clearpage

\section{NUMERICAL RESULTS}\label{sec:numerics}
\subsection{Numerical model using the octupole approximation}

In our numerical model we treat the gravitational effects of the
third body in the octupole approximation, meaning we average over the orbital
periods of both the inner binary and the outer companion and retain terms up to $(a_{in}/a_{out})$ to $3^{rd}$ order. Beside the perturbations due to the presence of the third body via Kozai--Lidov mechanism, we include
the following dynamical effects:
\begin{itemize}
\item periastron advance of the inner binary due to general relativity;
\item periastron advance arising from quadrupole distortions of the white dwarfs due to both tides and rotation;
\item orbital decay due to tidal dissipation in the white dwarfs;
\item loss of binary orbital angular momentum due to gravitational radiation.
\end{itemize}
\noindent The equations used in our model are those of \citet{2002ApJ...578..775B} for the octupole terms, which are based on those derived in \citet{2000ApJ...535..385F}, combined with equations from \citet{2012ApJ...747....4P} for tidal effects. Further discussion of the importance of octupole approximation can be found in \citet{2011Natur.473..187N, 2012arXiv1206.4316N, 2013MNRAS.431.2155N}. For the discussion of the relevance of the secular approximation we refer reader to Section \ref{sec:discussion}.  % and we list them in the appendix (to be added).

%\subsection{Results}

Following \citet{2011ApJ...741...82T}, as fiducial parameters we use: $m_1=0.8M_{\bigodot}$, $m_2=0.6M_{\bigodot}$,
and $m_3=1M_{\bigodot}$. The semimajor axis of the inner binary is
$a_{in}= 0.05 AU$ and the semimajor axis of the outer binary is $a_{out}=1AU$. We use the crude approximation that the radius of both white dwarfs in the inner binary is
$R= 5\times10^8\cm$, while the fiducial Love number is $k_2=0.1$ and tidal dissipation factor is $Q=10^7$. For the initial eccentricities and arguments of a periapse we take: $e_{in,0}=0.1$, $e_{out,0}=0.5$, $\omega_{in, 0}=\omega_{out,0}=0$ throughout this paper. Initially we take the WDs to be tidally locked. We only evolve systems within the following range of initial mutual inclinations: $85^{o}\leq i_{0} \leq102^{o}$. We consider system merged when $r_p\leq 2R_{sum}=4R$. Next we discuss the results of our model for:  high mutual inclination $91^{o}\leq i_0 \leq96^{o}$ and moderately high inclinations $85^{o}\leq i \leq90^{o}$ and $97^{o}\leq i_0 \leq102^{o}$.

\subsection{High mutual inclination $91^{o}\leq i_0 \leq96^{o}$}\label{sec:high}

\begin{figure}
\epsscale{1.0} 
\plotone{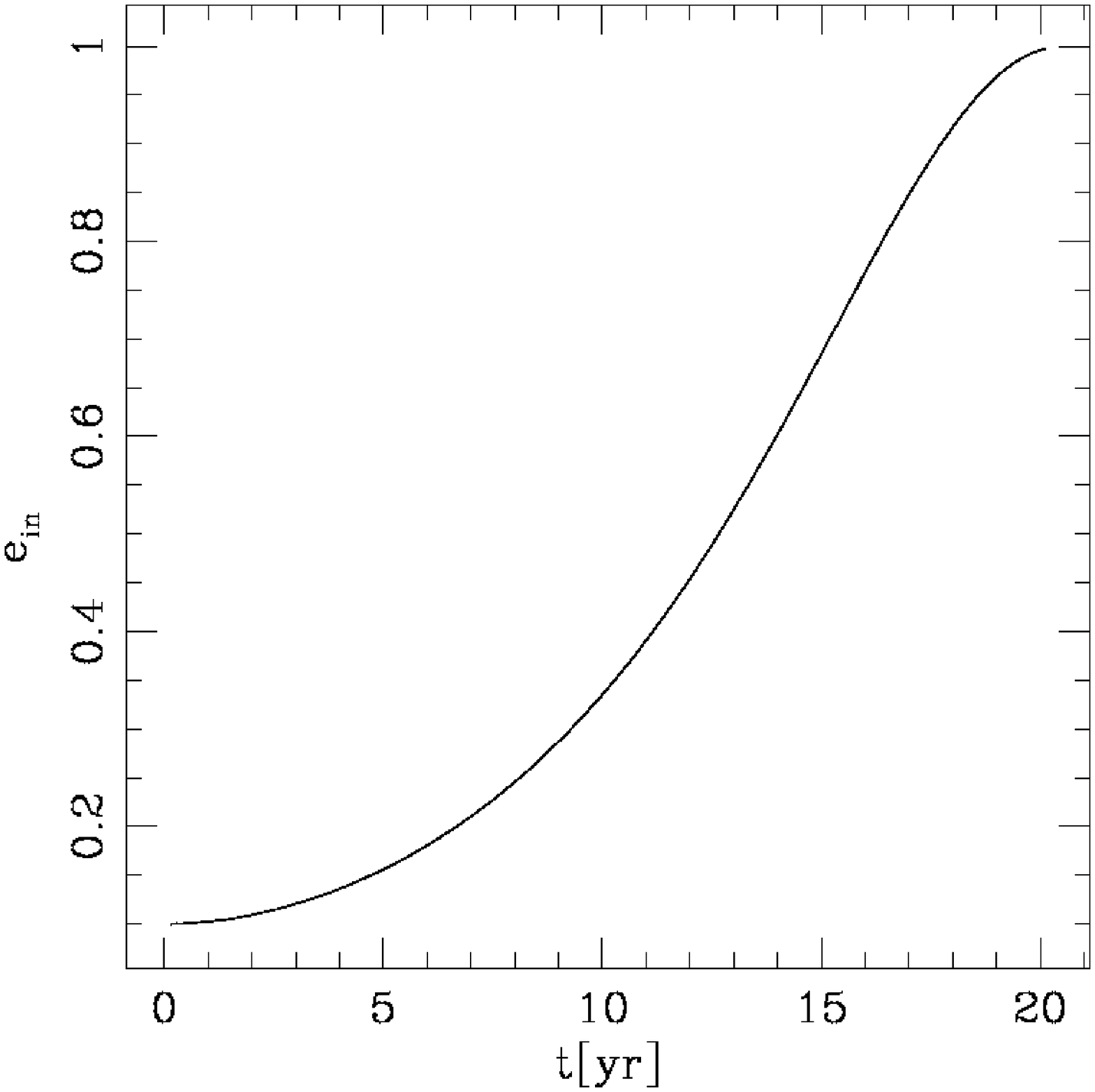}
\plotone{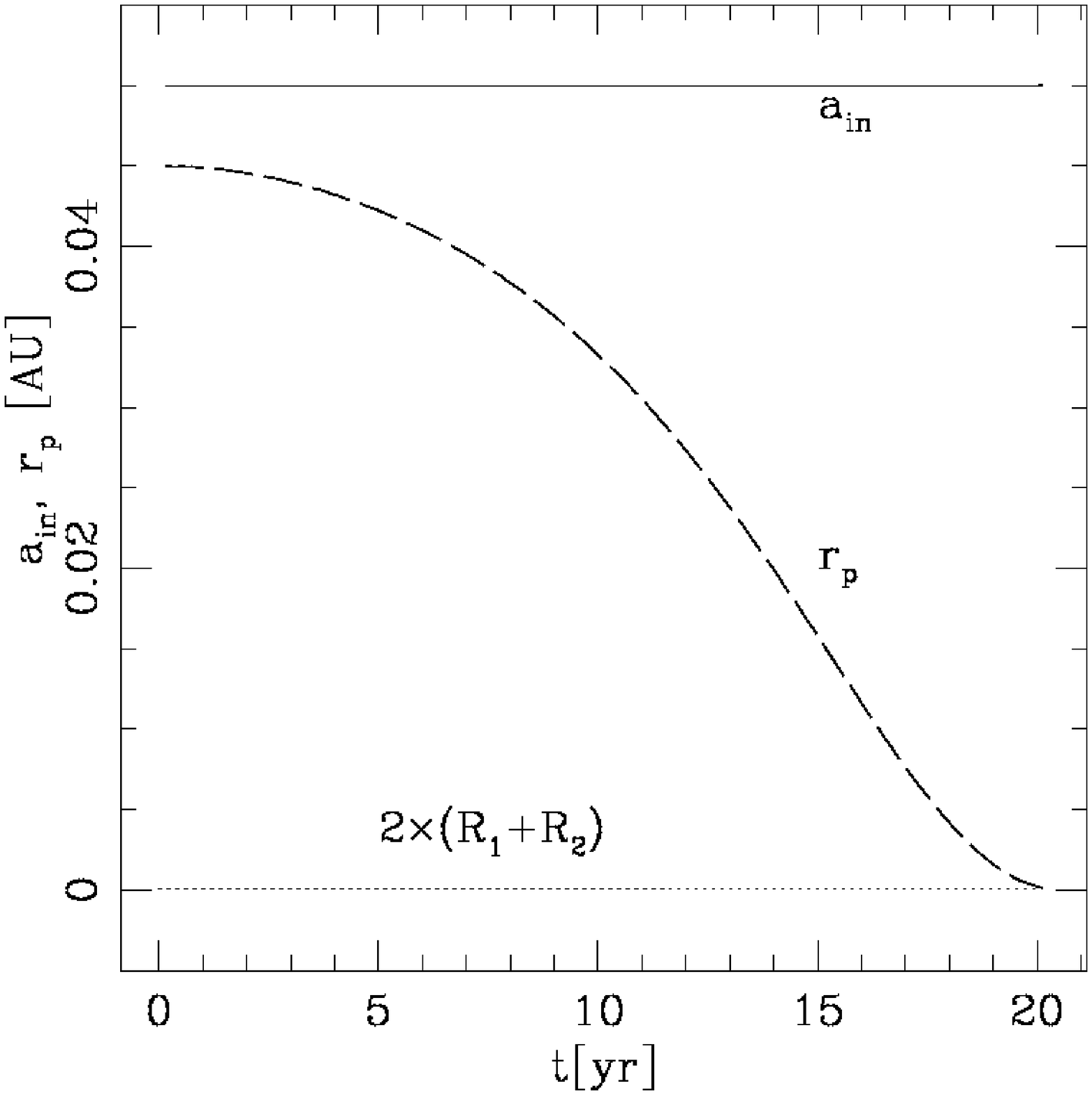}
\caption{
  The eccentricity as a function of time (upper
  panel) and the semimajor axis and the periapse of the inner binary as a function of time (lower panel) for our fiducial
  model. The solid line represents the semimajor axis while the dashed line shows the periapse evolution for $i_0=95^{o}$. The horizontal dotted line represents the double sum of the radii of the two white dwarfs. At the first eccentricity maximum the periapse is less than $4R=2\times10^9\cm$. We consider $r_p\leq4\times R$ as the condition for collision/merger in the inner binary. \label{Fig:ecc}}
\end{figure}

In this subsection we consider the parameter space where the initial mutual inclination is in the range $91^{o}\leq i_0 \leq96^{o}$. For such high initial mutual inclinations, as the eccentricity of the inner binary reaches its first maximum, the periapse of the inner binary, $r_p$, becomes less than $2\times (R_1+R_2)=4\times R$, and we consider that a collision/merger has occurred. Figure \ref{Fig:ecc}, upper panel, shows the evolution of the eccentricity of the inner binary. Within $20$ yr, the first eccentricity maximum occurs leading to the collision of the two white dwarfs. The lower panel of figure \ref{Fig:ecc} shows the evolution of the semimajor axis and the periapse of the inner binary. The semimajor axis remains constant while the periapse rapidly becomes of order of a few R. In this high inclination regime the sources of additional apsidal precession such as tidal and rotational bulges as well as tidal dissipation do not influence the outcome of the evolution. The time scale for the collision induced by the presence of a third body via Kozai oscillation is too short for any kind of tidal interaction to matter, which we confirm by repeating the calculation where we neglect tidal effects.  The possibility of Kozai cycle leading straight to a collision was noted in \citet{2011ApJ...741...82T} even though effects of rotational bulges and tidal dissipation were not taken into account. The range of initial mutual inclinations leading to direct collision becomes larger for wide range of the inner semimajor axes and moderately hierarchical triples when non secular approach is used,  as in work of \citet{KatzDong2012} (see Section \ref{sec:discussion} for detailed discussion). 
%
%%\clearpage
%%\clearpage
%

\subsection {Moderately high inclinations $85^{o}\leq i_0 \leq90^{o}$ and $97^{o}\leq i_0 \leq102^{o}$}\label{sec:moderate}

\begin{figure}
\epsscale{1.0} 
\plotone{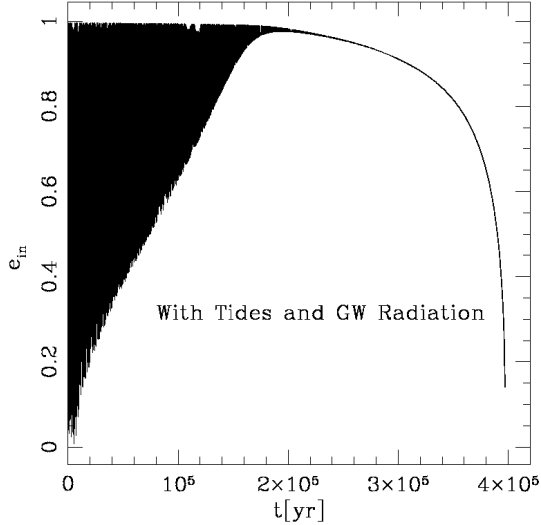}
\plotone{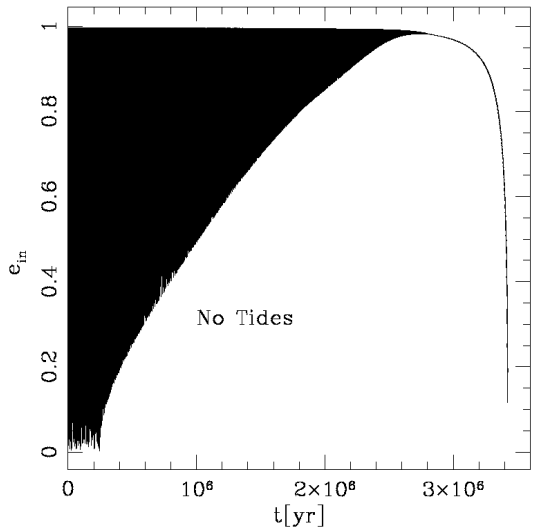}
\caption{
  The eccentricity as a function of time: upper panel shows the case where we take into account tidal effects and the lower panel shows the case where only GR precession and GW radiation are taken into account for our fiducial
  model with $i_0=89^{o}$.  The tidal dissipation factor is $Q=10^7$.  We terminate the integrations when $r_p\leq4R$. As the upper panel shows, including tidal dissipation leads to a merger timescale about an order of magnitude shorter than the merger time due to GW radiation alone. \label{Fig:e89}}
\end{figure}

\begin{figure}
\epsscale{1.0} 
\plotone{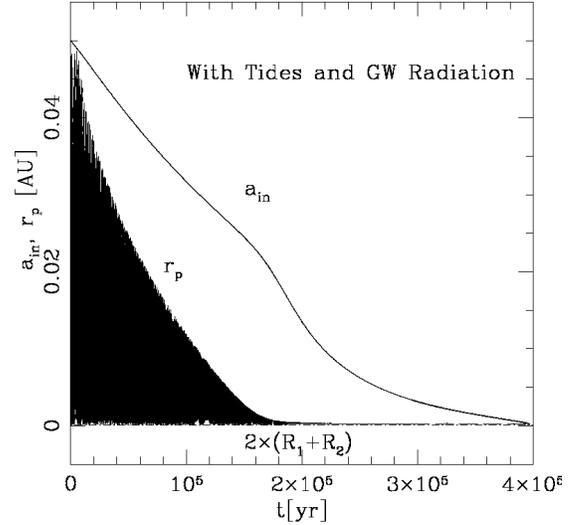}
\plotone{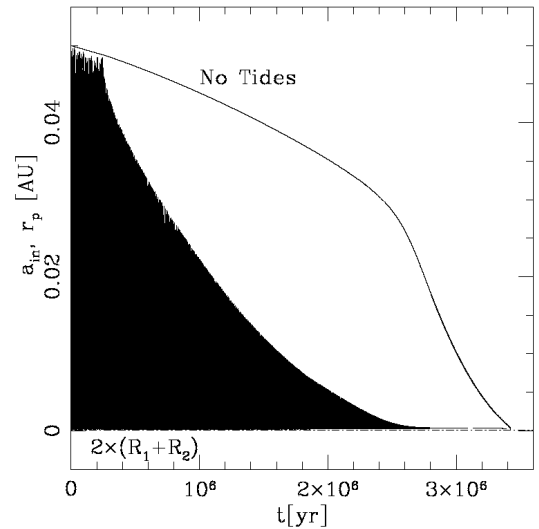}
\caption{
  The semimajor axis and the periapse of the inner binary as a function of time: upper panel shows the case where we take into account GR, GW and tidal effects, while the lower panel shows the case where only GR precession and GW radiation are taken into account. The plot represents our fiducial
  model, with $i_0=89^{o}$. The solid line represents the semimajor axis while the dashed line shows the periapse evolution. The dotted line represents the double sum of the radii of the two white dwarfs. As the upper panel shows, including tidal dissipation leads to a merger timescale about an order of magnitude shorter than the merger time due to GW radiation alone. \label{Fig:semi89}}
\end{figure}

In this subsection we explore the parameter space of moderately high initial mutual inclinations: $85^{o}\leq i_0 \leq90^{o}$ and $97^{o}\leq i_0 \leq102^{o}$. In the first case, calculations are done including all the dynamical effects captured in our model:  the secular perturbations due to the presence of the third body,  tidal effects such as tidal dissipation and tidal and rotational bulges, GR precession, and GW radiation. In the second case, we neglect tidal effects and we take into account only GR precession as a source of additional apsidal precession and GW radiation as the only source of dissipation in the system.  By design, the maximum eccentricity in the case of moderately high initial mutual inclinations is not high enough to lead to a collision. Instead, during the phases of high eccentricity the tidal dissipation and GW radiation are strong. As a consequence both the eccentricity and the semimajor axis of the inner binary decrease and eventually  the periapse becomes comparable to the sum of the radii of the two white dwarfs. Figures \ref{Fig:e89} and \ref{Fig:semi89} show the time evolution of the eccentricity, the semimajor axis and the periapse of the inner binary with and without tidal effects taken into account for $i_0=89^{o}$. The high eccentricity induced by the third body results in strong dissipation even with just GW radiation taken into account, which by itself shortens dramatically the merger timescale. However, comparing the merger timescale in the two cases clearly   demonstrates that the merger  timescale is at least an order of magnitude shorter when tidal dissipation is taken into account. 

%%\clearpage
%%\clearpage
%

%
%%\clearpage
%%\clearpage
%

%
%%\clearpage
\begin{figure}
\epsscale{1.0} 
\plotone{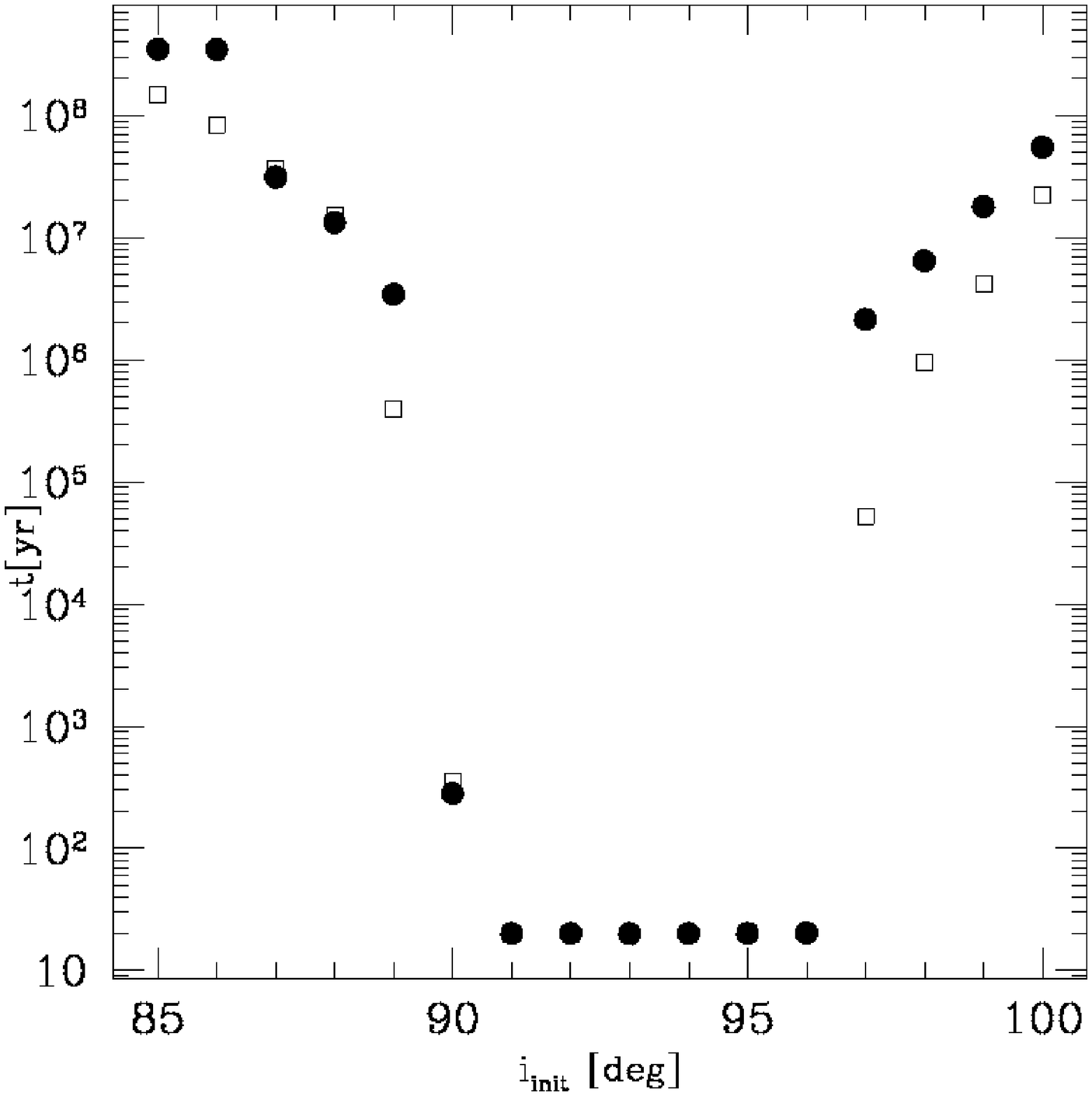}
%\plotone{semi89no_tides.eps}
\caption{
  The merger time as a function of initial inclination. The open squares come from numerical integration of the equations of motion that  include tidal dissipation and GW radiation while solid circles come from integration that includes only GW radiation. In the high inclination regime ($91^{o}\leq i_0 \leq96^{o}$), the collision of the two white dwarfs occurs at the first eccentricity maximum on a very short timescale ($\sim 20~yr$) showing that the suppressing effects due to additional sources of apsidal precession and dissipation are insignificant. Including tidal dissipation in the moderately high inclinations regime ($85^{o}\leq i_0 \leq90^{o}$ and $97^{o}\leq i_0 \leq102^{o}$)  leads to shorter merger timescale up to an order of magnitude comparing to the merger time for GW radiation alone. \label{Fig:time_merge}}
\end{figure}
%%\clearpage
%

Figure \ref{Fig:time_merge} shows the merger timescale dependence on the initial mutual inclination for the cases with and without the tidal effects included. In the high inclination regime ($91^{o}\leq i_0 \leq96^{o}$), the two white dwarfs collide at the first eccentricity maximum on a very short timescale ($\sim 20~yr$), so the suppressing effects due to additional sources of apsidal precession and dissipation are insignificant. On the other hand, in the moderately high inclination regime, tidal dissipation shortens the merger timescale for an additional order of magnitude as seen in figure  \ref{Fig:time_merge}. This implies that a combination of the perturbations due to the presence of a third body and tidal dissipation and GW radiation indeed dramatically shorten the merger timescale of WD--WD binaries in comparison to Hubble time, $T_{Hubble}$ when the system is in moderately high inclinations regime ($85^{o}\leq i_0 \leq90^{o}$ and $97^{o}\leq i_0 \leq102^{o}$). %Here we take $Q=10^7$, but one should keep in mind that smaller $Q$ leads to even faster tidal dissipation. 

\section{DISCUSSION}\label{sec:discussion}

In \citet{2011ApJ...741...82T}, the relevance of WD--WD, NS--WD, and NS--NS mergers driven by GW radiation  was explored for events like Ia supernovae, GRBs, and other transients. It was explicitly demonstrated that the GW merger timescale for these compact binaries becomes shorter by up to a few orders of magnitude due to the eccentricity oscillations induced by the presence of a third body at high inclination with the respect to the inner orbit, compared to the case of an isolated binary. The triple scenario allows for a wider range of binary semimajor axes leading to a merger in a Hubble time, enhancing the population of compact object binaries capable of producing Ia supernovae and GRBs. 

In this paper, we build on work of \citet{2011ApJ...741...82T} and explore the combined effect of tides and GW radiation on the merger timescale of WD--WD binaries in triple systems. We examine the evolution of the WD--WD binary in the presence of a third body at $a_{out}/a_{in}\sim 20$. When the mutual inclination is sufficiently high, the third body perturbs the inner binary orbit causing periodic oscillations in the eccentricity and mutual inclination via the Kozai--Lidov mechanism.  As discussed in Section \ref{sec:dynamics}, tidal effects become important during the phases of high eccentricity where the inner binary periapse is of order of few stellar radii. In the case of high mutual inclination ($91^{o}\leq i_0 \leq96^{o}$), the outcome of the evolution is a direct collision  of the WDs at the first eccentricity maximum, where the Kozai torque is the dominant torque and all other dynamical effects are insignificant. In other words, none of the additional sources of precession such as GR and tidal and/or rotational bulges have  timescale short enough to affect the maximum possible eccentricity or suppress Kozai cycles, contrary to what was expected in \citet{2011ApJ...741...82T}. For the same reason the strong tidal dissipation or GW radiation do not affect the evolution of the system even though the eccentricity reaches values close to $1$ as seen in Section \ref{sec:high}. Such collisions have been discussed by several authors as possible channels for production of type Ia supernovae \citep{2009ApJ...705L.128R, 2009MNRAS.399L.156R, 2010ApJ...724..111R, KatzDong2012, 2013MNRAS.430.2262H}. As shown here and in \citet{2011ApJ...741...82T}, this scenario seems very promising for retrograde orbits with $i_0 \geq 90^{o}$. Globular clusters, where the third star can be captured in binary-binary interactions \citep{2008msah.conf..101I, 2006MNRAS.372.1043I} may produce such retrograde triples. Further study of such triples formed via binary-binary interactions is a subject of our next paper. 

In \citet{KatzDong2012}, collisions of white dwarfs occur in moderately hierarchical triple systems ($3 \lesssim r_{p, out} /a_{in} \lesssim 10$) for wide range of $a_{in}$, where the inner binary reaches high eccentricities via Kozai--Lidov mechanism. A similar scenario is studied in \citet{2013MNRAS.430.2262H}. \citet{KatzDong2012} use a symplectic three body integrator while \citet{2013MNRAS.430.2262H} combine the secular three body dynamics with a detailed prescription for stellar, binary and triple evolution prior to the formation of the WD--WD binary. Considering the evolution on the main sequence is important because most of the systems that could be possible progenitors of WD--WD binaries (large $a_{in}$, small $a_{out}/a_{in}$) experience mergers while they are on main sequence \citep{2013MNRAS.430.2262H}. Therefore the rate of direct collisions described in \citet{KatzDong2012}, as well as in our work, contributes only a small fraction to the SNe Ia events in field. %The rate of direct collisions described in \citet{KatzDong2012} is not known because the distribution of inclinations after stellar evolution and mass loss to form a WD-WD binary is not known.   
 There is hope for getting this mechanism to work in globular clusters, where binary-binary interactions produce triples 
with a flat distribution in $\cos i$, but in the field, for a primordial triple, it is not clear what $\cos i_0$ should be once the stars have evolved to WDs.

\citet{KatzDong2012} find that the secular treatment of the Kozai--Lidov mechanism breaks down for most of their systems. The reason lies in the assumed ratio $R/a_{in}$, which for their choice of parameters requires $1-e_{in}< 10^{-6}$ in order for tides to be significant as for collisions to occur. This is not the case in our work since we are focused on very compact WD--WD binaries. For our choice of $R/a_{in}$, the two WDs collide when  $1-e_{in}\sim10^{-3}$. The secular treatment of the Kozai--Lidov mechanism gives an adequate description in this case (see equation 16 in \citet{KatzDong2012}).  

Our choice of parameter space is relevant for close compact object binaries in the field, as well as for those in globular clusters. For example, \citet{2009ApJ...705L.128R} and \citet{2009MNRAS.399L.156R} consider WDs that reside in the dense core of the globular clusters and similar dense stellar environments where the fact that stars are sufficiently close to each other makes collisions very likely. The authors propose WD--WD collision scenario with small impact parameter (close to head on collision) as a possible formation channel for type Ia supernovae in such environments. They carry out 3D hydrodynamical calculations of thermonuclear explosion of colliding WDs. \citet{2009ApJ...705L.128R} investigate the outcome of direct, head on collisions of  several mass pairs, while \citet{2009MNRAS.399L.156R} investigate the outcome of a collision of a single mass pair ($2 \times 0.6M_{\bigodot}$) with three different impact parameters. Both papers establish that a collision of moderately massive WD pair naturally  leads to shock-triggered ignition and a synthesis of substantial amounts of $Ni$ for small impact parameter even if the total mass of the pair is below the Chandrasekhar limit. As emphasized by both groups, one should keep in mind that the outcome of the WD--WD collision hinge on several factors: masses, nuclear compositions,  and impact parameters. Results from \citet{2009MNRAS.399L.156R} imply that the WD--WD collision with large impact parameters (i.e. grazing collision)  lacks violent shocks seen in the cases with small impact parameters. Instead, there is negligible nuclear burning as well as a negligible amount of $^{56}Ni$ produced by the initial interaction. In this scenario the WDs form a rotating disk of debris that cools down and eventually collapses into a single compact object.

 \citet{2012ApJ...747L..10P} study the violent merger of two massive white dwarfs. In their scenario the secondary (less massive white dwarf) eventually becomes dynamically unstable and is disrupted on a time scale of one orbit. The material from a disrupted white dwarf is violently accreted onto the surface of the  primary white dwarf where it gets compressed and heated up. Such a violent accretion leads to formation of the hot spots in which carbon burning is ignited. Once when the first hot spot reaches critical density and temperature, the detonation is assumed. Even though authors use high resolution SPH simulation they note that their work only suggests that detonation formation in the hot spot is plausible. 

\citet{2012MNRAS.422.2417D} study the outcome of white dwarf binary mergers by exploring the wide range of parameter space where they consider white dwarf masses between $0.2$ and $1.2M_{\bigodot}$ and their different chemical compositions. Their work demonstrates that large fraction of He-accreting binaries is  anticipated to explode prior to merger or at surface contact. All systems with primary mass under $1.1M_{\bigodot}$ experience such explosion as long as the He-donating white dwarf exceeds $\sim 0.4M_{\bigodot}$. Some similar systems could explode even earlier due to stream-induced detonations \citep[for details see][]{2010ApJ...709L..64G}. On the other hand, as \citet{2012MNRAS.422.2417D} show WD--WD binaries made entirely of carbon and oxygen are unlikely to explode prior or during the merger,  unless they contain sufficient amount of He in their outer layers \citep{2009ApJ...699.1365S,2012ApJ...746...62R}. \citet{2007MNRAS.380..933Y} showed that some of such systems may explode long after the dynamical merger, where characteristic time delay between dynamical merger and explosion is of order of $10^5\yr$.
The collisions presented in this paper correspond to either grazing collisions described in \citet{2009MNRAS.399L.156R}  or violent mergers described in \citet{2012ApJ...747L..10P}. The dynamical process leading to a merger/collissions in our study is relevant for cases where the donor is He white dwarf as well, and therefore can lead to explosions described in \citet{2012MNRAS.422.2417D}. The collisions described in \citet{KatzDong2012} correspond to head on collisions in both \citet{2009ApJ...705L.128R} and \citet{2009MNRAS.399L.156R}.  According to \citet{2009ApJ...705L.128R} and \citet{2009MNRAS.399L.156R}, the rates of collisions that may result in explosion are low and still subject to uncertainties (i.e. core-collapse evolution of the globular cluster). But, such rates indicate that WD--WD collision are not unlikely and hence they can contribute to type Ia events. The likelihood of small impact parameter collisions due to the eccentricity oscillations induced by the presence of a third body remains unknown. Therefore due to many uncertainties in this model one should be cautious about claiming that these events can explain the SNe Ia rates, but as already pointed out these events definitely do contribute to the overall rate.

%This may be excluded: In both studies the rates of collisions that may result in explosion are low and still subject to uncertainties (i.e. core-collapse evolution of the globular cluster). But, such rates indicate that WD--WD collision are not unlikely and hence they can contribute a modest fraction of type Ia events. The likelihood of small impact parameter collisions due to the eccentricity oscillations induced by the presence of a third body remains unknown. 

% In addition, Kushnir et al. 2013 argue that type Ia supernovae are indeed result of these head-on collisions of WDs in triple system. Using 2D hydrodynamical calculations, the authors showed that collision of two WDs with typical masses $0.5-0.9 M_{\bigodot}$ could lead to an explosion that synthesize $56Ni$ masses in the range of $0.15-0.8M_{\bigodot}$ corresponding to the distribution observed for the majority of the SNe Ia. 

In the regime of moderately high inclinations previously described additional sources of precession affect the evolution of the system by suppressing the Kozai--Lidov mechanism as demonstrated in Section \ref{sec:moderate}. We showed that including tidal dissipation together with GW radiation shortens the merger timescale by a factor of a few to an order of magnitude compared to when only GW radiation is taken into account (see Figure 4). The fact that tidal dissipation can speed up the mergers driven by GW radiation implies that the very prompt merger rate ($< 10^8$ yr) may be higher than found in \citet{2011ApJ...741...82T}. The tidal dissipation rate is determined by the tidal dissipation factor Q. In this work we take $Q\simeq10^7$ as obtained in \citet{2012ApJ...747....4P}, a value in a reasonable agreement with the findings of \citet{2011ApJ...740L..53P} and \citet{2011MNRAS.412.1331F}. %For the detailed discussion on rates, progenitors and gravity waves coming from akin triple systems we refer reader to \citet{2011ApJ...741...82T}.  

%
%%\clearpage
%\begin{figure}
%\epsscale{0.6} 
%\plotone{f3.eps}
%\plotone{f3_b.eps}
%\caption{
%  The eccentricity as a function of time (upper
%  panel) and the phase space ($e$ versus $\omega$) for our fiducial
%  model. The period of the eccentricity oscillations is $171$ days,
%  and the amplitude of the eccentricity oscillation is sufficient to
%  produce the observed factor of $2-3$ variation in luminosity.
%\label{Fig:ecc}}
%\end{figure}
%%\clearpage
%
%
%%
%\be %$
%\left({\dot P\over P}\right)_{obs} =
%\left({\dot P\over P}\right)_{Roche}+
%\left({\dot P\over P }\right)_{accel}+
%\left({\dot P\over P}\right)_{TD}.
%\ee %$
%%

\acknowledgments S.P. is grateful to Fabio Antonini, Smadar Naoz and Enrico Ramirez-Ruiz for useful comments and suggestions. This research has made use of the SIMBAD database, operated at CDS,
Strasbourg, France, and of NASA's Astrophysics Data System. The
authors are supported in part by the Canada Research Chair program and
by NSERC of Canada. T.A.T. is supported in part by NSF AST \#1313252.

\bibliography{1820}{}

\end{document}